\documentclass[aps,prl,a4paper,twocolumn,showemail]{revtex4}

\usepackage{graphicx}
\usepackage{bm}
\usepackage{amsmath}

\begin{document}

\title{Spin Squeezing of Atomic Ensembles via Nuclear-Electronic Spin Entanglement}

\author{T. Fernholz}
\author{H. Krauter}
\author{K. Jensen}
\author{J. F. Sherson}
\altaffiliation{Present address: Institut f\"ur Physik,
Johannes-Gutenberg Universit\"at, D-55099 Mainz, Germany}
\author{A. S. S$\o$rensen}
\author{E. S. Polzik}
\email{polzik@nbi.dk}
\affiliation{QUANTOP, Danish National
Research Foundation Center for Quantum Optics, Niels Bohr
Institute, Copenhagen University, DK 2100, Denmark}

\date{\today}

\maketitle



\textbf{Entangled many body systems have recently attracted
significant attention in various contexts. Among them, spin
squeezed atoms and ions have raised interest in the field of
precision measurements, as they allow to overcome quantum noise of
uncorrelated particles \cite{Kitagawa91, Leibfried, Geremia05,
Petersen, Roos, Windpassinger}. Precise quantum state engineering
is also required as a resource for quantum computation, and spin
squeezing can be used to create multi-partite entangled states
\cite{SoerensenPRL}. Two-mode spin squeezed systems
\cite{Julsgaard01} have been used for elementary quantum
communication protocols \cite{Julsgaard04, Sherson06}. Until now
spin squeezing has been always achieved via generation of
entanglement between different atoms of the ensemble \cite{Hald,
Kuzmich, Geremia}. In this Letter, we demonstrate for the first
time ensemble spin squeezing generated by engineering the quantum
state of each individual atom. More specifically, we entangle the
nuclear and electronic spins of $10^{12}$ Cesium atoms at room
temperature. We verify entanglement and ensemble spin squeezing by
performing quantum tomography on the atomic state.}

Ensembles of neutral atoms in the form of cold gases as well as
room-temperature vapor have emerged as important systems for the
generation of entanglement and for the storage of quantum
information \cite{Julsgaard04, Sherson06, Duan, Matsukevich, Chou,
Chen, Simon, Honda, Appel}. It has been shown that ensemble spin
squeezing can be used to significantly improve the fidelity of
some existing quantum memory protocols \cite{Julsgaard04}. In the
majority of the work on ensemble entanglement alkali atoms used as
memory units were treated as spin-$1/2$ systems, while their
actual much higher total angular momenta were considered an
unfortunate complication. Entanglement in those experiments could
therefore be generated only between different atoms of the
ensemble. In this Letter, we generate entanglement in an ensemble
of spin-$4$ atoms by using their internal structure to squeeze the
individual spins. Indeed, we can show that spin squeezing within
the Cesium $F=4$ manifold is a signature of entanglement between
the electron and the nucleus. However, as discussed below, the
ensemble spin squeezing does not necessarily follow from the
single atom spin squeezing. Therefore, we verify the collective
squeezing by measurement of the collective spin state via strong
coupling to off-resonant light. In particular, we perform for the
first time quantum tomography of a non-classical state of an
atomic ensemble.

We first consider a collection of ground state atoms, each
prepared in the magnetic sublevel $|F=4,m_F=4\rangle$ with respect
to the $x$ quantization axis. This fully stretched state is a
product state, where both the nuclear and the electronic spin are
in Eigenstates of their angular momentum projection operators with
maximal Eigenvalues $I_x=7/2$ and $S_x=1/2$, respectively. In the
following, we denote the total spin operator of the ensemble
$\mathbf{\hat{J}}$. For large atom numbers $N$, the total spin
becomes macroscopic with $\langle\mathbf{\hat{J}}\rangle=(J,0,0)$
in cartesian coordinates, and $J=\hbar N F$. The commutation
relation $[\hat{J}_y,\hat{J}_z]=i \hbar\hat{J}_x$ imposes an
uncertainty relation, and the fully pumped ensemble is in a
minimal uncertainty state with
$(\Delta\hat{J}_y)^2=(\Delta\hat{J}_z)^2=\hbar J/2$. Such states
are called coherent spin states (CSS) and set the standard quantum
limit (SQL) to spin projection measurements. For a squeezed spin
state (SSS), one of the orthogonal projections exhibits an
uncertainty below the SQL. For useful spin squeezing, however, it
is not sufficient to reduce the uncertainty below its initial
value ($\chi^2=2(\Delta\hat{J}_{\perp})^2/\hbar J<1$), but to
compare it to a CSS with the same mean spin ($\zeta^2=2(\Delta
\hat{J}_{\perp})^2/\hbar\langle|\mathbf{\hat{J}}|\rangle<1$)
\cite{Kitagawa91,Kitagawa}, or even more rigorously to the initial
uncertainty in spin angle ($\xi^2=2J(\Delta
\hat{J}_{\perp})^2/\hbar\langle|\mathbf{\hat{J}}|\rangle^2<1$)
\cite{Wineland92,Wineland94}. The last condition is the strongest
and provides a sufficient condition for entanglement of elementary
spin-$1/2$ constituents \cite{SoerensenNature}. It has to be met
to improve quantum metrology with a given number of atoms, e.g.,
the precision of atomic clocks.

Unconditional spin squeezing can be achieved by applying a
Hamiltonian that depends non-linearly on cartesian spin components
orthogonal to the mean spin \cite{Kitagawa}. A Hamiltonian of the
form $\hat{H}_s=\nolinebreak\alpha\hat{F}_z^2$ acting on a spin
$\mathbf{\hat{F}}$ induces a process called one-axis twisting.
Starting from a CSS aligned along the $x$-direction, the spin is
rotated about the $z$-axis in proportion to its fluctuating
$z$-component. The initially symmetric quasi-probability
distribution (QPD) will be sheared and appear squeezed in the $y,
z$ plane in a slightly rotated basis $\hat{F}_{y'},\hat{F}_{z'}$.
At the same time, the magnitude of the mean spin is reduced as the
QPD bends around the spherical phase space for constant $F$.

\begin{figure}
\includegraphics[width=85 mm]{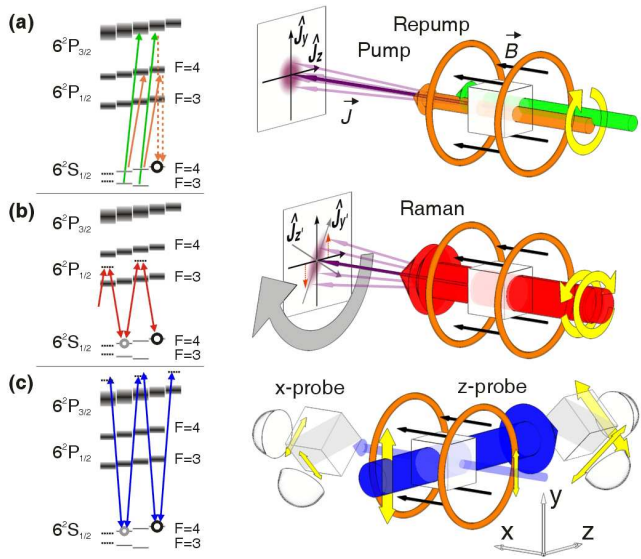}\\
\caption{Pulse sequence and level scheme for atomic state
preparation and quantum state reconstruction. (a) Optical pumping
creates a macroscopic spin $\vec{J}$ in a CSS, aligned with the
magnetic field $\vec{B}$. (b) A Raman transition creates a
coherent superposition of even magnetic sublevels, which shears
the QPD in the rotating frame and decreases the mean spin. (c) The
squeezed spin state of the ensemble is characterized via
off-resonant Faraday interaction with a probe beam, propagating in
the $z$-direction. Quantum noise limited polarization measurements
reveal the statistics of the atomic angular momentum projections
$\hat{J}_{y'}$ and $\hat{J}_{z'}$, which are precessing about the
$x$-axis. Simultaneously, a weak measurement along the $x$-axis
determines the magnitude of the mean atomic spin
$\langle\hat{J}_x\rangle$. See methods section for details.
\label{figSetup}}
\end{figure}

In our experiment we generate a squeezing Hamiltonian using the
tensor light shift induced by an off-resonant light field. In case
of linear polarization along the $z$-direction with classical
field amplitude $E_z$, the interaction with an atomic spin
$\mathbf{\hat{F}}$ reduces to an atomic Hamiltonian of the form:
\begin{equation}
H_s=-\frac{1}{4}(\alpha_0|E_z|^2+\alpha_2 |E_z|^2 \hat{F}_z^2).
\end{equation}
While the scalar polarizability $\alpha_0$ simply results in an
overall Stark shift, its tensor part, proportional to $\alpha_2$,
provides the basis for spin squeezing.

In the $x$-basis, the $z$-polarized field can be decomposed into
equally strong $\sigma_+$ and $\sigma_-$ components, driving Raman
transitions between magnetic sublevels ($\Delta m=2$). In the
presence of a magnetic field along the $x$-direction, the above
description remains valid after transformation to a frame
co-rotating with the Larmor frequency $\omega_L$. Consequently,
the frequencies of the two Raman fields have to be shifted by
$\pm\omega_L$ to meet the two-photon resonance condition (see
Fig.~\ref{figSetup}(b)). In this sense, spin squeezing corresponds
to the creation of paired excitations of the spin at frequency
$2\omega_L$ and will reveal itself in the Fourier component of the
rotating spin at frequency $\omega_L$.

Naturally, this type of interaction is non-linear only in spin
components of individual atoms. Therefore, the maximally possible
degree of squeezing \cite{SoerensenPRL} is limited by the total
spin of a single atom, while the total spin of an ensemble can be
arbitrarily high. With one-axis twisting, the squeezing is further
limited by the deviation of the resulting QPD from a geodesic in
the spherical phase space. This finding is equivalent to the
differential light shift imposed on magnetic sublevels in the
$x$-basis, which inevitably accompanies the Raman coupling and
detunes the coupling between different sublevels. A somewhat
higher degree of squeezing can be achieved with a two-axis
countertwisting Hamiltonian of the form
$H_s=\alpha(\hat{F}_z^2-\hat{F}_y^2) $\cite{Kitagawa}. Taking the
decreasing mean spin into account, squeezing parameters of
$\chi^2\approx0.163$, $\zeta^2\approx0.247$, and
$\xi^2\approx0.327$ can be reached with this Hamiltonian, using
the maximum available spin of $F=4$ in ground state Cesium atoms.
The required Hamiltonian could, e.g., be implemented with two
laser fields of orthogonal polarizations (along rotating $y'$- and
$z'$-directions) and opposite detunings from an atomic resonance.
Here, we balance the light shift instead with the second-order
Zeeman shift $\hat{H}_Z=\beta\hat{F}_x^2$, such that all energy
splittings between magnetic substates of the $F=4$-manifold
coincide. By adjusting the light intensity and detuning,
$\alpha_2|E_z|^2=-8\beta$ can be chosen to obtain:
$\hat{H}_s+\hat{H}_Z=-\frac{\alpha_0|E_z|^2}{4}+\beta \hbar^2
F(F+1)/2+\beta(\hat{F}_z^2-\hat{F}_y^2)/2, $ using
$\hat{F}_x^2+\hat{F}_y^2+\hat{F}_z^2=\hbar^2 F(F+1)$ for
Eigenstates of $\mathbf{\hat{F}}^2$.

Coherent superpositions of magnetic sublevels similar to those
described above have been recently generated by universal quantum
control of a hyperfine spin in an ultra-cold ensemble
\cite{Chaudury}. In that experiment the mean values of elements of
a single atom density matrix were determined. However, knowledge
of the single atom density matrix is not sufficient to infer spin
squeezing of the ensemble. Due to the collective preparation, the
members of the ensemble will not in general be uncorrelated.
Collective spin squeezing, as required for quantum limited
metrology or quantum memory \cite{Julsgaard01}, can therefore only
be determined via measurement of collective quantum fluctuations,
ideally via full quantum tomography, as performed in the present
paper. Only such measurements can reveal the deleterious effect of
classical and quantum atom-atom correlations, which can be created
in the preparation process and can easily ruin the spin squeezing
of the ensemble. In particular, it is of major importance to send
the driving Raman field (Fig.~\ref{figSetup}(b)) along the
direction of the mean spin ($x$-direction). If, e.g., the
orthogonal $y$ direction is used, the quantum fluctuations of the
driving field will couple via atomic $\pi$-transitions to the
collective symmetric mode for $\Delta m_x=1$ coherences, and thus
destroy collective spin squeezing under strong coupling
conditions. In fact, this process forms the basis for quantum
memory applications \cite{Julsgaard04,Sherson06}.

The protocol for atomic quantum tomography used to reconstruct the
collective quantum state is detailed elsewhere
\cite{Sherson06,Sherson,Usami}. In brief, we send a strong,
linearly $y$-polarized beam along the $z$-axis of the atomic
ensemble, see Fig.~\ref{figSetup}(c). After the interaction, we
perform polarization homodyning in the $45^\circ$-basis, measuring
the photon number difference in the two outputs, i.e. the Stokes
operator $\hat{S}_y=\hat{n}_{+45^\circ}-\hat{n}_{-45^\circ}$. Due
to the Faraday effect, the polarization of the input beam is
rotated proportionally to the instantaneous atomic spin component
$\hat{J}_z$, producing a proportional homodyne output for small
rotations. The atomic spin precesses, and information on the two
orthogonal, rotating spin components $\hat{J}_{y'}$ and
$\hat{J}_{z'}$ can be obtained by evaluating the cosine and sine
components of the output signal at the Larmor precession frequency
$\omega_L$. The atomic ensemble is freshly prepared and the
procedure repeated $10^4$ times to determine the quantum
statistics including the mean and variance of the collective
atomic spin state.

\begin{figure}
\includegraphics[width=85 mm]{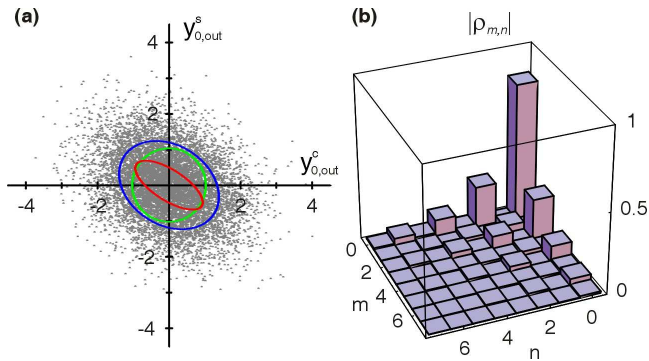}\\
\caption{Exemplified quantum state reconstruction (at $\approx
3$~dB squeezing) with $\kappa^2\approx 0.8$. (a) Scatter plot of
$10^4$ realizations. The scaled atomic covariance (red ellipse
with $\sqrt{2}\sigma$-radii) is determined by correcting the total
covariance (blue) for light and back-action noise (green). (b)
Maximum likelyhood estimation of the density matrix $\hat{\rho}$
for the collective spin state, showing alternating excitation
numbers $m,n$ in the harmonic oscillator approximation. See
methods section. \label{figReconstruction}}
\end{figure}

\begin{figure}
\includegraphics[width=85 mm]{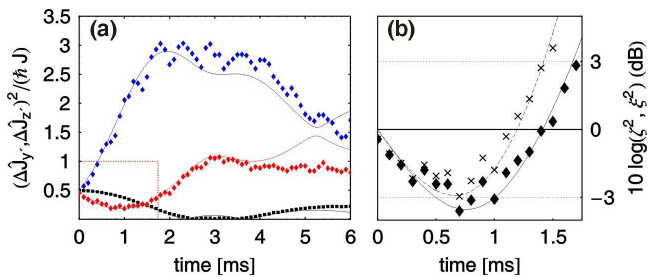}\\
\caption{(a) Results of the quantum state reconstruction for
varied Raman pulse length. Red and blue diamonds show atomic
squeezed ($\equiv\chi^2/2$) and anti-squeezed variances,
respectively. For useful squeezing, they have to be compared to
the variance of a CSS with the same mean spin (black squares). (b)
Resulting squeezing parameters $\zeta^2$ (diamonds) and $\xi^2$
(crosses), representing renormalized data from the indicated
region in (a). Theoretical predictions are shown as solid and
dashed lines.\label{figResults}}
\end{figure}

The state tomography is illustrated in
Fig.~\ref{figReconstruction}. Atomic variances are directly
inferred from the sampled data. In addition, we performed maximum
likelyhood estimations of the collective state in the 10
best-resolved dimensions of Hilbert space \cite{Hradil}. The
results, i.e. the measured variances, are shown in
Fig.~\ref{figResults} together with predictions from a density
matrix calculation. Our model includes atomic decay due to
spontaneous emission and incorporates independently measured
values for depolarizing and dephasing times in the dark
($T_1=80$~ms, $T_2=20$~ms). We kept the effective light power in
the cell as a free model parameter, but especially for short
times, the theoretical results are stable against small variations
in the control parameters. The used light power slightly
overcompensates the second-order Zeeman-shift, and our model
suggests better performance at even higher powers as it identifies
atomic decay as the dominant limitation. We consistently achieve
squeezing of a canonical variable by $\zeta^2=0.47\pm0.1$. Quantum
memory \cite{Julsgaard04} and metrology applications gain by
$\xi^2=0.54\pm0.1$.

In conclusion, we demonstrated unconditional squeezing of a
collective spin with a noise reduction of $\approx-3dB$ by
generation of entanglement within individual members of the
ensemble. We have fully characterized the non-classical collective
atomic state via quantum state tomography. The method developed in
this paper is independent and complementary to the ensemble
entanglement and squeezing approach based on
quantum-non-demolition (QND) measurements
\cite{Julsgaard01,Kuzmich,Geremia}. Indeed it can, e.g., be
directly applied, either separately or in combination with the QND
method, to enhance the fidelity of the quantum memory
\cite{Julsgaard04}. It may also find use in spectroscopic
applications, such as magnetometry beyond the SQL.

\section{Methods}
\subsection{State preparation}

In our experiment, we employ a room temperature ensemble of
$N\approx 10^{12}$ $^{133}$Cs atoms, contained in a paraffin
coated, $22\times 22\times 22$~mm$^3$ glass cell. The cell is
shielded against field fluctuations and subjected to a homogeneous
magnetic field along the $x$-direction of $B_x\approx 0.9$~G. By
fine-tuning the magnetic field we adjust the atomic Larmor
precession to a reference frame rotating at $\omega_L=2\pi\times
322$~kHz, which is defined by a computer controlled radio
frequency (rf) synthesizer.

The experimental pulse sequence consists of three stages and is
depicted in Fig.~\ref{figSetup}. The two-step state preparation
starts with an 8~ms long optical pump pulse of two resonant,
circularly polarized beams, propagating along the $x$-direction.
See Fig.~\ref{figSetup}(a). After optical pumping, we find 98\% of
the atoms in the $|F=4,m_x=4\rangle$ state, by evaluating the
magneto-optical resonance signal (MORS) \cite{Julsgaard}. At the
second stage, we coherently transfer atomic population between
even magnetic substates of the $F=4$-manifold by driving Raman
transitions on the $D_1$-line for a variable duration $T_R=0-6$~ms
with a single photon detuning of $\Delta_R=-550$~MHz from the
$6^2S_{1/2},F=4\rightarrow6^2P_{1/2},F'=4$ transition. See
Fig.~\ref{figSetup}(b). The two-photon difference $\nu_R\approx
644.2$~kHz is resonant with the $m_x=4\rightarrow m_x=2$
transition, including corrections for second-order Zeeman and
differential light shifts. Finally, the collective atomic state is
analyzed via off-resonant Faraday interaction with a strong
($P\approx 5$~mW), linearly polarized, 2~ms long probe pulse. The
probe laser frequency is close to the $D_2$-line with a
blue-detuning of $\Delta_P=825$~MHz from the $F=4\rightarrow F'=5$
transition. See Fig.~\ref{figSetup}(c).

Some experimental issues have to be considered for the generation
of collective squeezed states. The state preparation must be
completed on a timescale short compared to decoherence mechanisms,
e.g. caused by inhomogeneous fields, atom-atom, and atom-wall
collisions. Therefore, the strength of the squeezing interaction
should be maximized. Although atomic motion leads to averaging, it
is necessary, particularly on short time scales, to provide rather
homogeneous coupling strength over the cell volume with expanded,
mode-matched Raman fields. A small angle between the two beams
leads to a varying relative phase over the cell volume, which
gives rise to a residual Doppler shift for moving atoms and
results in incoherent squeezing. A similar effect is caused by the
collective spin of the polarized ensemble. Due to the
Faraday-effect, the sample is circularly birefringent and induces
a phase shift over the cell length, $\Delta\phi<12^\circ$ for our
parameters. A simple and robust experimental approach is to
generate the necessary Raman fields by amplitude modulation of a
single, linearly polarized light beam. But to avoid the additional
sidebands and minimize spontaneous emission, we mode-match two
phase-stable, circularly polarized fields. For ground state alkali
atoms, the tensor polarizability compared to the scattering rate
is limited by the excited state hyperfine splitting. Their ratio
is optimal when tuned between the hyperfine transitions of the
$D_1$ line, ($F=4\rightarrow F'=3,4$) at 894~nm for Cesium. At
this detuning, the tensor polarizability is also largest and
stationary and thus independent of atomic velocity. In addition,
it has the correct sign for compensation of the second-order
Zeeman shift.

\subsection{State reconstruction}
The commutation relation prevents us from measuring both spin
components with arbitrary precision, and their precise values are
masked by back-action of the original light noise. Their
statistical moments, however, can be inferred from the moments of
the light output. For large atom numbers, we can make a harmonic
oscillator approximation and define canonical operators as
$\hat{x}=\hat{J}_{y'}/\sqrt{\hbar\langle\hat{J}_x\rangle}$,
$\hat{p}=\hat{J}_{z'}/\sqrt{\hbar\langle\hat{J}_x\rangle}$ for
atomic variables and
$\hat{y}=\hat{S}_{y}/\sqrt{\langle\hat{S}_x\rangle}$,
$\hat{q}=\hat{S}_{z}/\sqrt{\langle\hat{S}_x\rangle}$ for light
variables of a given mode with
$[\hat{x},\hat{p}]\approx[\hat{y},\hat{q}]\approx i$. The two
simultaneously accessible output variances are given by:
\begin{equation}
(\Delta \hat{y}^{\textrm{c},\textrm{s}}_{0,\textrm{out}})^2 =
(\Delta \hat{y}_0^{\textrm{c},\textrm{s}})^2 +\frac{\kappa^2}{2}
(\Delta\hat{x},\Delta\hat{p})^2 +\frac{\kappa^4}{12} (\Delta
\hat{q}_1^{\textrm{s},\textrm{c}})^2,
\end{equation}
where the indices correspond to different, non-orthogonal temporal
modes of the light field, oscillating as cosine (c) and sine (s)
components at the Larmor frequency. Index $0$ refers to a
rectangular envelope of duration $T$, and $\kappa^2\propto
J_x\,T$.

We scale the time-integrated homodyne output by referencing it to
the light vacuum noise, $\langle(\Delta\hat{y})^2\rangle=
\langle(\Delta\hat{q})^2\rangle=1/2$. To avoid changes in the
optical and electronic paths, this is measured by shifting the
atomic Larmor frequency well outside the detection bandwidth using
the magnetic field. The coupling strength $\kappa^2$ is calibrated
by measuring the resonant noise of the atomic ensemble in thermal
equilibrium. For the thermal state, the back-action noise is zero
and the output noise is given by the light noise and the atomic
contributions from thermally populated states in the
$F=4$-manifold: $(\Delta
\hat{y}^{\textrm{c},\textrm{s}}_{0,\textrm{out}})^2
=\frac{1}{2}+\frac{\kappa^2}{2 \hbar J_x}\,(\Delta
\hat{J}_{y',z'})^2=\frac{1}{2}+\frac{\kappa^2 N
\hbar}{2\,J_x}\cdot\frac{15}{16}$. In principle, the second-order
Zeeman-shift of the magnetic sub-levels has to be considered, but
it is compensated by the differential light shift induced by the
probe laser. The measurement of the mean spin $J_x/N$ is
calibrated with a fully pumped ensemble.

The density matrix for the collective spin state is represented in
a basis given by excitation operators defined as
$b,b^{\dagger}=(\hat{x}\pm i\hat{p})/\sqrt{2}$ and
$\hat{m},\hat{n}=b^{\dagger}b$. See Fig.~\ref{figReconstruction}.

\section{Acknowledgements}
We thank J. H. M\"uller and M. Christandl for helpful discussions.
This work was supported by the EU under contracts FP6-015848 (QAP)
and FP6-511004 (COVAQUIAL).

\end{document}